\pdfoutput=1
\documentclass[12pt]{iopart}
\usepackage{dcolumn}
\usepackage[T1]{fontenc}
\usepackage[latin9]{inputenc}
\usepackage{graphicx}
\usepackage[letterpaper]{geometry}
\usepackage{wrapfig}
\usepackage{graphics}
\usepackage{color}
\usepackage{epsfig}
\usepackage{units}
\usepackage{dsfont}
\usepackage{circle}
\usepackage[svgnames]{xcolor}
\usepackage{stmaryrd}

\expandafter\let\csname equation*\endcsname\relax
\expandafter\let\csname endequation*\endcsname\relax

\newcommand{\fig}[1]{Fig.~\ref{#1}}

\newcommand*{\greysquare}{\textcolor{gray}{\blacksquare}}
\newcommand*{\whitesquare}{\textcolor{white}{\square}}

\usepackage{amsmath}
\usepackage{amssymb}

\begin{document}

\title{Topological transition in disordered planar matching: combinatorial arcs expansion}

\author{Andrey Y. Lokhov$^{1}$, Olga V. Valba$^{2,3}$, Sergei K. Nechaev$^{1,3,4}$, and Mikhail V. Tamm$^{3,5}$ }
\address{
$^{1}$Universit\'e Paris-Sud/CNRS, LPTMS, UMR8626, B\^at. 100, 91405 Orsay, France,\\
$^{2}$N.N. Semenov Institute of Chemical Physics of the Russian Academy of Sciences, 119991, Moscow, Russia,\\
$^{3}$Department of Applied Mathematics, National Research University Higher School of Economics, 101000, Moscow, Russia,\\
$^{4}$P.N. Lebedev
Physical Institute of the Russian Academy of Sciences, 119991, Moscow, Russia,\\
$^{5}$Physics Department, Moscow State University, 119991, Moscow, Russia.\\
}
\ead{andrey.lokhov@lptms.u-psud.fr, valbaolga@gmail.com, sergei.nechaev@lptms.u-psud.fr, thumm.m@gmail.com}

\begin{abstract}
In this paper, we investigate analytically the properties of the disordered Bernoulli model of
planar matching. This model is characterized by a topological phase transition, yielding complete
planar matching solutions only above a critical density threshold. We develop a combinatorial
procedure of arcs expansion that explicitly takes into account the contribution of short arcs, and
allows to obtain an accurate analytical estimation of the critical value by reducing the global
constrained problem to a set of local ones. As an application to a toy representation of the RNA
secondary structures, we suggest generalized models that incorporate a one-to-one correspondence
between the contact matrix and the RNA-type sequence, thus giving sense to the notion of effective
non-integer alphabets.
\end{abstract}


\maketitle

\section*{Introduction}

The problem of optimal matching in a given set of interacting variables under specific non-local
topological constraints in presence of quenched disorder is a challenging question in information
theory, statistical physics, and biophysics. A particularly important case of such a global
constraint is given by the requirement for the optimal matching
configurations to have a planar structure. Indeed, the planar diagrams play a key role in many areas, including matrix and
gauge theories \cite{Brezin1978}, many-body condensed matter physics \cite{AbrikosovGorkov1975},
quantum spin chains \cite{Saito1990}, random matrix theory \cite{Mehta2004}. Another area where
planar matching appears naturally is the biophysics of secondary structures of RNA molecules
\cite{DeGennes1968,Nussinov1980,Mueller2003,BundschuhHwa2002}: the RNA molecules differ from other
biologically active associating polymers, for instance proteins, by a formation of hierarchical
``cactus-like'' secondary structures, topologically isomorphic to a tree. In other words, the bonds
between monomers can be drawn in a form of a planar diagram with non-intersecting arcs, while the
configurations that do not obey this property are suppressed \cite{vanBatenburg2000}.

Matching problems have attracted considerable attention in mathematics, physics and computer science communities \cite{Lov\'asz1986}. An equivalent dimer covering problem on planar lattices has been studied by Kasteleyn \cite{Kasteleyn1961}.
Recently, the existence of a new phase transition has been reported in the problem of complete
planar matching on a line \cite{ValbaTammNechaev2012,LVTN}. For a sequence of $L$ points, an instance of the problem is given by a symmetric $L \times L$ contact matrix $A$ with entries $A_{ij}$ taking values one (if matching between $i$ and $j$ is in principle possible) and zero (if a link $(i,j)$ is forbidden). The question that we are trying to answer is whether it is possible to draw a complete matching of $L/2$ \emph{non-intersecting} arcs involving all the points (see \fig{fig:dyckpath}~(a) for an example). In \cite{LVTN}, we assumed that the entries of the contact matrix are generated according to the simple Bernoulli model, i.e. they are independently equal to one with probability $p$, and are equal to zero otherwise. We have shown that perfect planar matching solutions follow a critical behavior: they exist only above a certain critical density of possible contacts, or unity entries in the contact matrix. Along with an accurate numerical study, two analytical estimations of the critical point
have been provided, however, making use of uncontrollable, to a certain extent, approximations. The
difficulty in these calculations arise essentially from the \emph{quenched} nature of the disorder
in the random contact matrix. One of the estimations featured the matrix model formulation
suggested in \cite{VernizziOrlandZee2005}, leading to a field theory with a complicated interaction
in a form of infinite series that needs to be averaged out. Another was based on a combinatorial
approach, explicitly accounting for the contribution of shortest arcs, observed to play the
dominant role in the studied planar structures. A quantification of this contribution provided a
good estimate for the transition point, however making use of a non-exact mean-field-like averaging
argument.

In this paper we go further in the last direction, developing a procedure for a detailed treatment
of quenched disorder at the level of shortest arcs in the complete planar matching problem. In
particular, we show how to get successive estimations to the value of the transition point via arcs
expansion, explicitly calculating the contributions of shortest and next-to-shortest arcs, and
treating the contribution of the rest in a mean-field manner. These calculations involve a
representation in terms of spin chain models, as well as combinatorial and generating functional
formalism. Aiming at the application for the random RNA-type sequences, we introduce two new models involving
explicit representation of an instance of the problem as a string of letters, and numerically study
the phase transition of interest.

The paper is organized as follows. In section 1, we provide definition of the model and briefly
report the previously established results. In section 2, we present an exact treatment of the first
order in arcs expansion, mapping the problem to a spin chain model and calculating the contribution
of the shortest arcs. In section 3, we show how to generalize these computations to include the
correlations arising from the next-to-shortest arcs in presence of a quenched disorder. Finally, in
the section 4 we numerically study other models, allowing for an explicit representation in form of
strings of letters, and investigate the effect of transitivity on the existence of the
perfect-imperfect phase transition, making connections to the fluctuations-free Bernoulli limit.

\section{Background and definitions}
\label{sec:background}

In this section we provide the definition of the model and recall some previously established results.

\subsection{Bernoulli model and mapping to Dyck paths}

\begin{figure}[ht]
\centering
\includegraphics[scale=0.48]{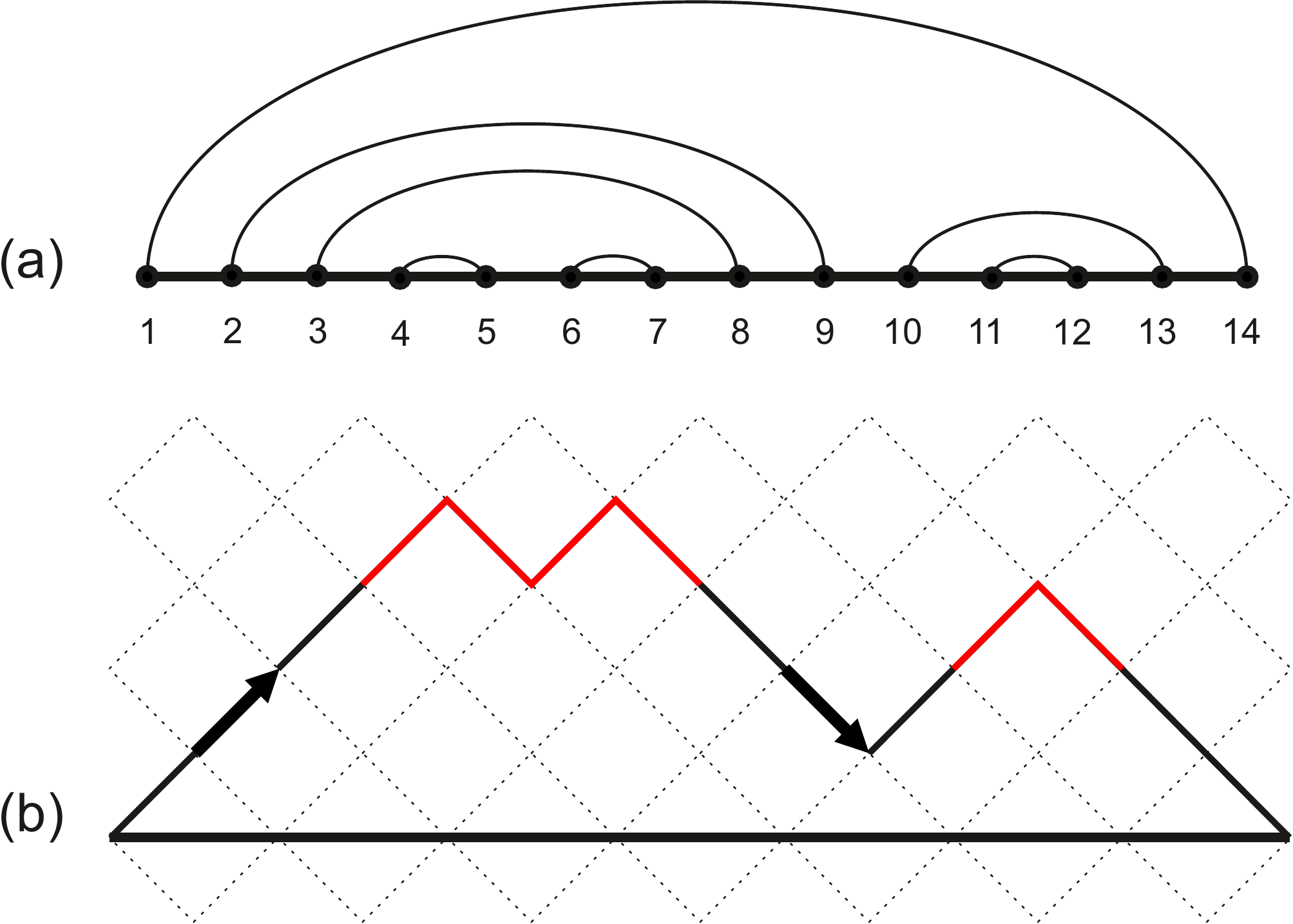}
\caption{Example of (a) a perfect planar matching configuration, and (b) the corresponding mapping
to a Dyck path. The arc is given by an ``up'' and ``down'' steps at the same height, shown by
arrows $\nearrow$ and $\searrow$. The part of the walk between arrows is a Dyck path itself. The
shortest arcs correspond to the peaks of the Dyck path representation and are marked with red.}
\label{fig:dyckpath}
\end{figure}

The Bernoulli model of complete planar matching is stated as follows. An instance of the problem is
given by a symmetric $L\times L$ random matrix $A$ containing zeros and ones. The upper-diagonal
entries $A_{ij}$ of this matrix ($i>j$) are independent identically distributed random variables,
generated by the distribution
\begin{equation}
{\rm Prob}(A_{ij})=p \delta(A_{ij}-1) + (1-p) \delta(A_{ij}),
\label{eq:Bernoulli_matrix_probability}
\end{equation}
where $\delta(x)=1$ for $x=0$, and $\delta(x)=0$ otherwise. In other words, each element
$A_{ij}=A_{ji}$ is independently either one with the probability $p$ for any $i\neq j$, or zero
otherwise. Now we take $L$ points $i=1, \ldots, L$ on a line, and draw $L/2$ non-intersecting arcs
between pairs of points allowed by the non-zero entries $A_{ij}$ such that each point is involved
in one link only and the links form a planar diagram, see~\fig{fig:dyckpath}(a). If at least one
such set of links exists, we say that the problem allows for the \emph{complete matching} solution.

The phase transition \cite{LVTN} in this problem occurs as the parameter of the model, $p$, reaches
a certain critical value of bond formation probability $p_{c}$. It can be equivalently thought of
as a transition in a constrained satisfaction problem \cite{Friedgut1999}: as the number of
constraints per node, imposed by the matrix $A$, is below a certain critical value, the problem
exhibits a complete matching solution, while otherwise no complete matching solution exists in the
limit $L \to \infty$.

In what follows, we use an important one-to-one mapping between complete $L$-point planar diagrams
and the $L$-step Brownian excursions, known as Dyck paths \cite{Lando2003}. In this representation,
also called mountain (or height) diagram, each monomers is represented by either an ``up''-step
($\nearrow$) or a ``down''-step ($\searrow$) with ``up''-steps corresponding to opening arcs, and
``down''-step to closing ones. An example is given in the \fig{fig:dyckpath}, with the steps up and
down at positions 2 and 9, corresponding to the arc between points 2 and 9 in the planar matching
structure. The total number of Dyck paths of even length $L$ is given by a Catalan
number
\begin{equation} C_{L/2}=\frac{L!}{(\frac{L}{2})!(\frac{L}{2}+1)!} \sim \frac{2^{L}}{L^{3/2}}
\sqrt{\frac{2^3}{\pi}},
\label{catalan}
\end{equation}
where the asymptotic expression is valid for $L \gg 1$. If $p=1$ in our matching problem, all the
planar configurations are solutions to the perfect matching problem, and their total number is then
also given by \eqref{catalan}. For $p<1$, the number of possible planar configurations is reduced,
and drops down to zero below a certain value $p_{c}$. A naive estimation of $p_{c}$ can be readily
obtained using the following mean-field-like argument. Since each arc in the diagram is present
with the probability $p$, the probability that the whole configuration is allowed, is given by
$p^{L/2}$. Assuming the planar diagrams in the ensemble of all possible ones are
\emph{statistically independent}, we get the probability $\mathcal{P}$ to have at least one perfect
planar matching configuration:
\begin{equation}
\mathcal{P}=1-(1-p^{L/2})^{C_{L/2}}=1-\exp\left(-p^{L/2}C_{L/2}\right),
\end{equation}
where the last equality is valid for $L \to \infty$, leading to the probability one for $p>p_{c}$,
and to the probability zero for $p<p_{c}$. The naive mean-field critical threshold $p_{c}$ is thus
given by the condition
\begin{equation}
\lim_{L \to \infty} p_c \left[C_{L/2}\right]^{2/L} = 1,
\label{eq:naive_mean_field}
\end{equation}
yielding $p_c=1/4$. However, here we have neglected the statistical correlations between different
possible configurations. For instance, let $\tau$ and $\rho$ be the two arbitrarily chosen
configuration of arcs out of $C_{L/2}$ possible configurations. The probability that they both
satisfy the constraints imposed by the contact matrix $A$ is given by $p^{L/2}p^{L/2}p^{-n_{\tau
\cap \rho}L/2}$, where $n_{\tau \cap \rho}$ is a fraction of common arcs in the configurations
$\tau$ and $\rho$. Therefore, equation \eqref{eq:naive_mean_field} provides only a crude estimation
to the true value of $p_c$, and it has to be generalized to
\begin{equation}
\lim_{L \to \infty} \xi(p_c) \left[C_{L/2}\right]^{2/L} = 1, \; \xi(p_c) = 1/4,
\label{eq:critcond}
\end{equation}
where $\xi(p)$ is some weight (due to correlations) that has to be determined. In the sections
\ref{sec:first-order} and \ref{sec:second-order}, we will see how to calculate the transition value
analytically in a more accurate way; before proceeding to the calculations, we present the related
numerical results.

\subsection{Numerical results}

Let us briefly discuss the numerical results obtained in \cite{LVTN}. Finding a maximum matching on
a graph is a problem of a polynomial complexity \cite{Micali1980}. Numerically, the phase
transition in the planar matching on a line can be identified using the dynamic programming
algorithm, for details see \cite{LVTN,ValbaTammNechaev2011}. The idea is that the ground state free
energy of the system, $F_{1,N}$, proportional to the number of nodes involved in the planar
matching (and equal to $L/2$ if the complete matching solution exists), can be computed iteratively
as a zero-temperature limit of the corresponding equation for the partition function
\cite{DeGennes1968,Nussinov1980}, using the following expression:
\begin{equation}
F_{i,i+k}=\max_{s=i+1,...,i+k} \left[F_{i+1,i+k}, A_{i,s}+F_{i+1,s-1}+F_{s+1,i+k}\right].
\end{equation}

\begin{figure}[ht]
\centering
\includegraphics[scale=0.39]{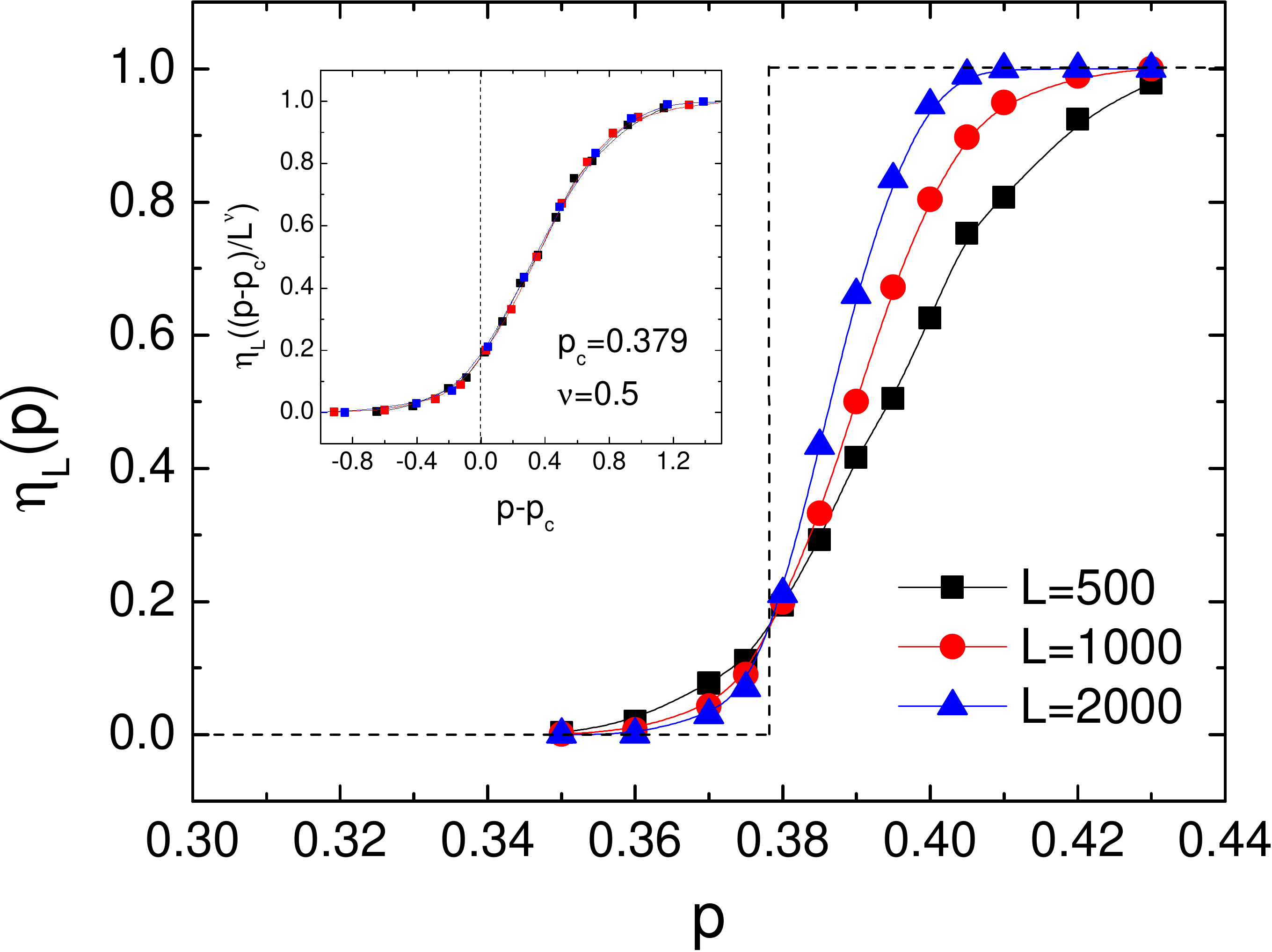}
\caption{Main figure: The fraction of perfect matchings $\eta_{L}(p)$ as a function of the density
$p$ of ones in the contact matrix $A$ for different lengths. The dashed line corresponds to the
phase transition in the thermodynamic limit $L\to\infty$ at the critical point $p_{c}=0.379$.
Inset: Finite-size scaling analysis of curves, corresponding to different lengths $L$. The fitting
procedure yields $\nu \approx 0.5$ as the value of the transition width exponent. Each data point
is averaged over $10^{4}$ instances.}
\label{fig:numtrans}
\end{figure}

Looking for the fraction, $\eta_{L}(p)$, of adjacency matrices, which allow for perfect planar
matchings, in the whole ensemble of random Bernoulli matrices, one has $\eta_{\infty}(p)=1$ for
$p>p_{c}$, and $\eta_{\infty}(p)=0$ for $p<p_{c}$. The finite-size results are shown in the
\fig{fig:numtrans} for different lengths, $L=500,\,1000,\,2000$. The phase $p>p_{c}$ corresponds to
a gapless complete matching, while in the phase $p<p_{c}$ the best possible matching always
contains a finite fraction of gaps. The scaling analysis places the phase transition point at $p_c
\approx 0.379$, and allows to estimate the power-law decay of the transition width $L^{-\nu}$, with
$\nu=0.5$.

\section{Improved analytical estimation of $p_{c}$ via arcs expansion: first order}
\label{sec:first-order}

In this section, we show how to obtain a refined estimation of the perfect-imperfect transition
point, using the formulation of the problem in terms of Dyck paths and combining exact
combinatorial and mean-field techniques. The method is based on the observation that the arcs with
smaller lengths are more likely to appear in the complete matching structure than those with higher
lengths. Indeed, recall that locally, in the complete matching configuration, the arc opened at $i$
and closed at $j$ corresponds to the part of a Dyck path, starting by an ``up''-step $\nearrow$ in
position $i$ and ending by the first ``down''-step $\searrow$ at the same height in position $j$,
cf. \fig{fig:dyckpath}(b). Hence, this random walk between $i$ and $j$ is a Dyck path itself, and
the probability to find an arc connecting $i$ and $j$, reads
\begin{equation}
P(i,j)=\frac{C_{(k-1)/2}}{2^{k+1}},
\label{eq:probarc}
\end{equation}
where $k=j-i$; the nominator represents the total number of Dyck paths of length $k$, given by
\eqref{catalan}, and the denominator is the total number of possible random walks of this length.

From \eqref{eq:probarc} we see that short links play an exceptional role in the formation of planar
configurations: $P(i,j)$ is non-zero for odd $k$ only, and few first values are: $P(i,i+1)=1/4$,
$P(i,i+3)=1/16$, $P(i,i+5)=1/32$, \emph{etc}. In particular, in the large $L$ limit, about a half
of all $L/2$ arcs are the shortest ones (``$S$-arc'') of length two (corresponding to red peaks in
the \fig{fig:dyckpath}(b)).

In our previous work \cite{LVTN} we have used this fact to provide an estimate for the
perfect-imperfect transition point by considering the following approximation:
\begin{equation}
\xi^{L/2}(p)= \underbrace{p^{L/4}}_{\rm long\;arcs}\, \underbrace{{\cal P}^{(1)}_{S}(p)}_{\rm S-arcs},
\label{eq:first_order_xi}
\end{equation}
that is, explicitly accounting for the correlations coming from shortest arcs, and assuming that
all longer arcs give a mean-field contribution $p^{L/4}$. Thus, the problem is reduced to placing
$L/4$ shortest arcs on the line of $L$ points, representing positions $(i,i+1)$, each position
being allowed or forbidden as dictated by the contact matrix values $A_{i,i+1}$. Note that since
the arcs can not share the same node, the $S$-arcs can not occupy neighboring positions $(i,i+1)$
and $(i+1,i+2)$ in such a placement.

We assume here that short arcs are uncorrelated apart from the non-overlap constraints. In real arc
structures it is not the case: indeed, the total number of available structures with exactly $k$
shortest arcs in the absence of disorder is known to be given by the so-called Narayana number
$N(2L,k)$ \cite{Narayana,Narayana2} instead of $C^{L/4}_{3L/4}$ (see the computation below).
However, correlations between short arcs are induced by the positions of longer ones, so assuming
them to be uncorrelated seems to be a natural first approximation, while the correlations will
arise naturally as one takes into account arcs of length 3, 5, \emph{etc}. (see section
\ref{sec:second-order} for more details).

We can express ${\cal P}^{(1)}_{S}(p)$ as follows:
\begin{equation}
{\cal P}^{(1)}_{S}(p)=\frac{B^{(1)}_{S}(p)}{B^{(1)}_{S}(1)},
\label{eq:first_order_bp}
\end{equation}
where $B^{(1)}_{S}(p)$ is the number of ways to put uniformly $L/4$ $S$-arcs, allowed by the
contact matrix $A$ of density $p$, on a line of $L$ points, according to the non-touching
constraint. It is easy to see that in the fully-connected case $p=1$ (all the positions are
allowed), $B^{(1)}_{S}(1)=C^{L/4}_{3L/4}$, corresponding to the placement of $L/4$ objects among
$L/4$ $S$-arcs (that we will denote by $\greysquare$) and $L/2$ unmatched vertices (denoted by
$\Circle$): the link configuration has then a form $(\cdots\Circle\Circle\Circle
\greysquare\greysquare\Circle\Circle\Circle \Circle\greysquare\Circle\greysquare\cdots)$. In
\cite{LVTN}, we have used the approximation $B^{(1)}_{S}(p)=C^{L/4}_{3pL/4}$, assuming that each
position in the ``circle-and-square'' representation is allowed with probability $p$. Strictly
speaking, it is not true, and gives only an upper-bound on the value of critical point computed at
this level. Indeed, the density of "1" on  the diagonal $A_{i,i+1}$ is equal to $p$ in the
thermodynamic limit, but the "1" are distributed independently, meaning that they may correspond to
incompatible neighboring positions $(i,i+1)$ and $(i+1,i+2)$; at the same time, the
``circle-and-square'' representation automatically incorporates the non-touching constraint. In
this section we derive a contribution of the $S$-arcs (or peaks in the Dyck path representation,
see \fig{fig:dyckpath}), via an exact procedure.

\subsection{Formulation of the problem as a spin chain model}

The problem of placing $L/4$ $S$-arcs on a line of $L$ a priori available positions $(i,i+1)$ can
be equivalently formulated as a diluted Ising spin chain model. To each couple $(i,i+1)$ we
associate a variable $\sigma_{i}$, equal to one if the arc is placed at this position, and and to
zero otherwise. Because of the non-touching constraint, the product $\sigma_{i}\sigma_{i+1}$ must
be always equal to zero. Moreover, we cannot count an arc as placed if the corresponding position
is forbidden by the contact matrix, i.e. if $q_{i}\equiv A_{i,i+1}=0$. In what follows, we will
denote allowed position ($q_{i}=1$) by a square with a dot $(i,i+1)\equiv\boxdot$, and the
forbidden position ($q_{i}=0$) by an empty square $(i,i+1)\equiv \square$. The number of placements
verifying these conditions are counted using the partition function
\begin{equation}
Z=\sum_{\{\sigma_{i}\}}e^{-\beta H[q,\sigma]}
\end{equation}
in the limit $\beta \to \infty$, where $H[q,\sigma]$ is given by
\begin{equation}
H[q,\sigma]=\sum_{i=1}^{L}q_{i}q_{i+1}\sigma_{i}\sigma_{i+1}
\end{equation}
with a counting constraint
\begin{equation}
\sum_{i=1}^{L}q_{i}\sigma_{i}=\frac{L}{4}\equiv M.
\end{equation}
Therefore, $Z$ can be expressed as
\begin{equation}
Z=\sum_{\{\sigma_{i}\}}e^{-\beta \sum_{i=1}^{L}q_{i}q_{i+1}\sigma_{i}\sigma_{i+1}}\frac{1}{2\pi
i}\oint \mu^{\sum_{i=1}^{L}q_{i}\sigma_{i}-(M+1)}d\mu=\frac{1}{2\pi i}\oint
\mu^{-(M+1)}Z_{\mu}d\mu.
\label{eq:zviazmu}
\end{equation}
Under periodic boundary conditions, $Z_{\mu}$ can be computed via the transfer matrix method:
\begin{equation}
Z_{\mu}=\sum_{\{\sigma_{i}\}}e^{-\beta \sum_{i=1}^{L}q_{i}q_{i+1}\sigma_{i}\sigma_{i+1}+\frac{1}{2}
\log{\mu}\sum_{i=1}^{L}(q_{i}\sigma_{i}+q_{i+1}\sigma_{i+1})}=\text{Tr}\prod_{i=1}^{L} T_{i,i+1},
\label{eq:producttm}
\end{equation}
where the transfer matrix $T$ reads
\begin{equation}
T_{i,i+1}=\begin{pmatrix}
e^{-\beta q_{i}q_{i+1}}\mu^{(q_{i}+q_{i+1})/2} & \mu^{q_{i+1}/2}\\
\mu^{q_{i}/2} & 1
\end{pmatrix}
\label{eq:transfermatrix1}
\end{equation}
The solution is easy to obtain explicitly in the fully-connected case $p=1$, when $q_{i}=1$ for all
$i$, and the chain has a form $(\boxdot\boxdot\cdots\boxdot)$. In this case, we have (in the limit
$\beta \to \infty$)
\begin{equation}
Z_{\mu}=\lambda_{1}^{L}+\lambda_{2}^{L},
\end{equation}
where $\lambda_{1,2}=(1\pm\sqrt{1+4\mu})/2$ are the eigenvalues of the matrix
\eqref{eq:transfermatrix1}. We get
\begin{equation}
Z_{\mu}=\frac{2}{2^{L}}\sum_{k=0}^{L/2}C_{L}^{2k}(1+4\mu)^{k}=
\frac{1}{2^{L-1}}\sum_{k=0}^{L/2}\sum_{l=0}^{k}C_{L}^{2k}C_{k}^{l}(4\mu)^{l}.
\end{equation}
From \eqref{eq:zviazmu}, $Z=Z[M,L]$ is non-zero only for $l=M$, and finally we get (using
combinatorial summation formula \cite{Prudnikov1986})
\begin{equation}
Z[M,L]=\frac{1}{2^{L-1}}\sum_{k=0}^{L/2}C_{L}^{2k}C_{k}^{M}(2)^{2M}=\frac{L}{M}C_{L-M-1}^{M-1}.
\label{eq:cisperiodic}
\end{equation}
In general, $p\neq 1$ case, some of the $L$ possible positions for the placement of the length-2
arcs are forbidden by the contact matrix. In other words, the contact matrix partitions the
length-$L$ chain of all possible shortest arcs positions into pieces, representing the sequences of
allowed positions, surrounded by forbidden ones:
$(\cdots\square\boxdot\boxdot\square\square\boxdot\square\boxdot\boxdot\square\cdots)$. Moreover,
we see that the sequence of forbidden positions of arbitrary length is equivalent to a single
forbidden position in the product \eqref{eq:producttm} up to a normalization constant,
$(\cdots\square\boxdot\boxdot\square\square\boxdot\square\boxdot\boxdot\square\cdots) \Rightarrow
(\cdots\square\boxdot\boxdot\square\boxdot\square\boxdot\boxdot\square\cdots)$: in this case the
transfer matrix \eqref{eq:transfermatrix1} reduces to
\begin{equation}
T^{0}_{i,i+1}=\begin{pmatrix}
1 & 1\\
1 & 1
\end{pmatrix},
\label{eq:transfermatrix2}
\end{equation}
and we have $(T^{0})^2=2T^{0}$. Let us denote $q_{k}$ the density (in the large $L$ limit) of
sequences of allowed positions of length $k$:
$\square\underbrace{\boxdot\boxdot\cdots\boxdot}_{k}\square$. We have
\begin{equation}
q_{k}=p^{k}(1-p)^2,
\end{equation}
where the two factors $(1-p)$ come from the forbidden positions nearest to the first and to the
last positions of the sequence, and each factor $p$ is the probability of an allowed position. It
is easy to check that the overall density of allowed positions for the shortest arcs must be equal
to $p$:
\begin{equation}
\sum_{k}kq_{k}=p(1-p)^2(1+2p+3p^2+\ldots)=p.
\end{equation}
Giving physical meaning to $B^{(1)}_{S}(p)$, we need to solve the following \emph{constrained
independent set (CIS) problem}: count the number of ways to distribute $L/4$ arcs such that they do
not touch each other, in the ensemble of allowed partitions. For each sequence it means that if a
certain position is chosen, other arcs cannot be placed in the neighboring positions, even if these
last are allowed by the contact matrix $A$. Note, however, that this global CIS problem is reduced
to a set of \emph{local} ones on the sequences with densities $q_{k}$: since they are separated by
at least one forbidden position, the distribution of $S$-arcs happens independently on each
sequence.

\subsection{Strict bound}

Let us first ask a simpler question: what is the \emph{maximum} number of arcs that can be placed,
given the densities $q_{k}$? It is easy to see that for a piece of length $k$, at most
$[(k+1)/2]\equiv r_{k}$ positions can be occupied under non-touching constraint. Therefore, the
maximum fraction of shortest arcs is
\begin{equation}
\sum_{k}\left[\frac{k+1}{2}\right]q_{k}=p(1+p)(1-p)^2(1+2p^2+3p^4+4p^6+\ldots)=\frac{p}{1+p}.
\end{equation}
As a by-product, we get a non-trivial strict bound on the value of $p_{c}$: since we need to place
at least $L/4$ arcs, we immediately conclude that $p>1/3$. It coincides with the lower bound for
RNA-type matching, found in \cite{Vladimirov2013} using the explicit construction in terms of
integer-valued alphabets.

\subsection{Solution of the CIS problem}

Now let us return to the solution of the local CIS problem as it has been stated previously. Let us
call $R_{m,k}$ the number of ways to put $m$ $S$-arcs on the allowed sequence of positions of
length $k$. For deriving this quantity, it is sufficient to notice that when one arc is placed, it
is no longer possible to place another arc on a neighboring position due to the non-touching
constraint. Hence, starting to put the $S$-arcs (represented by $\greysquare$) one by one on the
sequence of length $k$, we should forbid the position next to the placement position (this
constrained position will be denoted by $\boxtimes$):
\begin{equation}
\underbrace{\boxdot\greysquare\boxtimes\boxdot\boxdot\greysquare\boxtimes\cdots\boxdot}_{k}.
\end{equation}
In other words, we have to count the number of ways to put $m$ objects ($\greysquare\boxtimes$) on
a sequence of length $k$. This number is given by $C^{m}_{k-2m+m}$ if the last position of the
sequence is left free ($\boxdot$), and to $C^{m-1}_{(k-1)-2(m-1)+(m-1)}$, if it is occupied
($\greysquare$):
\begin{equation}
R_{m,k}=C^{m}_{k-m}+C^{m-1}_{k-m}=C^{m}_{k-m+1}.
\label{eq:cisseq}
\end{equation}
This result is in perfect agreement with the expression for the fully-connected case, when all the
$L$ positions are allowed: $B^{(1)}_{S}(1)=R_{L/4,L}=C^{L/4}_{3L/4}$. Notice that \eqref{eq:cisseq}
is different from the formula \eqref{eq:cisperiodic} (with obvious correspondences $M\to m$ and
$L\to k$), because the last one, being derived with periodic boundary conditions, corresponds to a
ring of allowed positions, rather than to a sequence. Anyway, one of these expressions can be
easily derived from another by noticing that a CIS problem on the sequence is equivalent to a CIS
problem on a ring with one additional forbidden position, leading to the equality
\begin{equation}
R_{m,k}+R_{m-1,k-2}=Z[m,k+1],
\end{equation}
which is verified, given \eqref{eq:cisperiodic} and \eqref{eq:cisseq}.

Given the solution of the local CIS problem \eqref{eq:cisseq}, it is easy to construct the solution
of the global problem. Let us introduce a generating function for a piece of length $k$:
\begin{equation}
Q_{k}(s)=\sum_{m=0}^{r_{k}}R_{m,k}s^{m}.
\end{equation}
Hence the generating function for the whole chain of $L$ a priori available positions for the
$S$-arcs reads
\begin{equation}
Q(s)=\prod_{k=1}^{L}(Q_{k}(s))^{Lq_{k}},
\end{equation}
or, explicitly,
\begin{equation}
Q(s)=(1+s)^{Lp(1-p)^2}(1+2s)^{Lp^2(1-p)^2}(1+3s+s^2)^{Lp^3(1-p)^2} (1+4s+3s^2)^{Lp^4(1-p)^2}\ldots
\end{equation}

\begin{figure}[!ht]
\begin{center}
\includegraphics[width=0.57\textwidth]{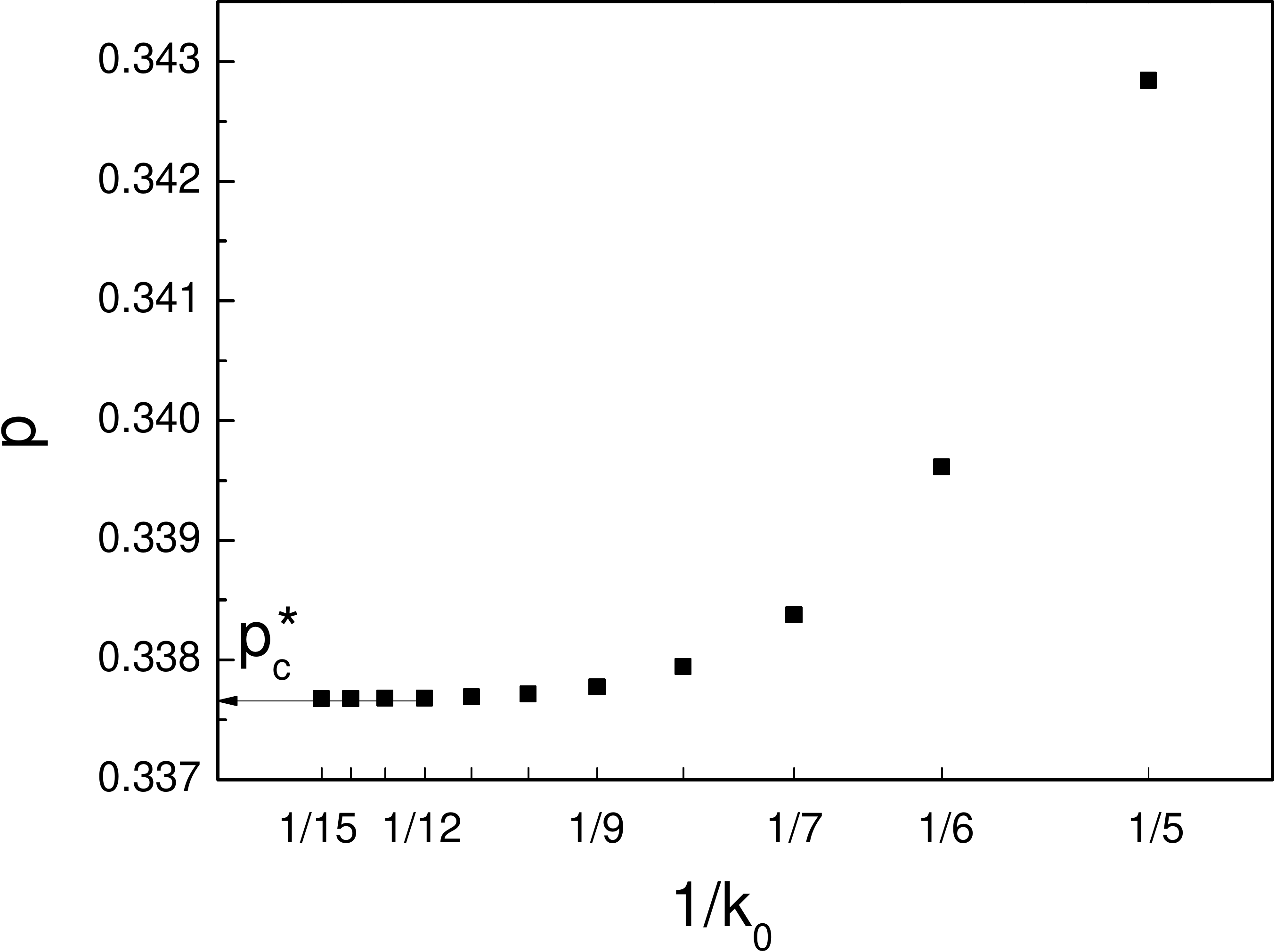}
\end{center}
\caption{Estimation of the prediction $p^{*}_{c}$ for the critical value $p_{c}$ at the first order of
arcs expansion. Each point represents a corresponding estimation when \eqref{eq:large_deviation} is
approximated by a partial sum up to the lengths $k_{0}$. The estimations demonstrate a fast
convergence with $k_{0}$ to the value $p^{*}_{c}=0.3376$.}
\label{fig:first_order}
\end{figure}

Since we want to place $L/4$ shortest arcs, we are interested in the coefficient behind the
$s^{L/4}$: this is exactly the quantity $B^{(1)}_{S}(p)$. This coefficient is given by the
integration of $Q(s)/s^{L/4}$ around zero:
\begin{equation}
B^{(1)}_{S}(p)=\frac{1}{2\pi i} \oint ds \exp \left[L((1-p)^2 f_{s}(p)-1/4 \log s) \right],
\label{eq:coefficient_integral}
\end{equation}
where
\begin{equation}
f_{s}(p)=\sum_{k=1}^{L}p^{k}\log \left( \sum_{m=0}^{[(k+1)/2]}C^{m}_{k-m+1}s^{m} \right).
\label{eq:large_deviation}
\end{equation}
Using explicit summation \cite{Prudnikov1986}, this result can be written as
\begin{equation}
f_{s}(p)=\sum_{k=1}^{L}p^{k}\log \left( \frac{(1+\sqrt{1+4s})^{k+2}-(1-\sqrt{1+4s})^{k+2}}
{2^{k+2}\sqrt{1+4s}} \right).
\label{eq:large_deviation_ressumed}
\end{equation}
Each term in this sum is decreasing, so in numerical calculations we can approximate this function
by partial sums to some order $k_{0}$. The integral \eqref{eq:coefficient_integral} can be treated
by the steepest descent method. The saddle-point equation reads
\begin{equation}
(1-p)^2\frac{\partial f_{s}(p)}{\partial s}=\frac{1}{4s}.
\label{eq:saddle_point}
\end{equation}
Given the solution $s^{*}$ of the equation \eqref{eq:saddle_point}, one gets the expression for
$B^{(1)}_{S}(p)$:
\begin{equation}
B^{(1)}_{S}(p)=\exp \left[ L( (1-p)^2 f_{s^{*}}(p)-1/4 \log s^{*}) \right].
\end{equation}
Approximating the large deviation function \eqref{eq:large_deviation} by partial sums up to the
fifteenth order, and combining with \eqref{eq:first_order_bp}, \eqref{eq:first_order_xi} and
\eqref{eq:critcond}, we get a fast convergence to the prediction of the critical point
$p^{*}_{c}=0.3376$, see \fig{fig:first_order}, providing an expected shift to a lower value from
the result $p^{*}_{c}=0.35$, originally found at this level of expansion in \cite{LVTN}.

\section{Improved analytical estimation of $p_{c}$ via arcs expansion: beyond the first order}
\label{sec:second-order}

The estimation obtained in the previous section can be systematically improved by considering
correlations arising form higher-length arcs. The idea is to use not only the information from the
diagonal $A_{i,i+1}$ of the contact matrix, but also from the diagonal $A_{i,i+3}$, i.e. to take
into account the constraints on the placement of $S$-arcs (shortest, length-2 arcs), that come from
the placement of the length-4, or next-to-shortest ($NS$) arcs. Therefore, we can write, as
previously
\begin{equation}
\xi^{L/2}(p)= \underbrace{p^{L/4}}_{\rm longer\;arcs}\,
\underbrace{{\cal P}^{(2)}_{S}(p)}_{\rm S-arcs},
\label{eq:second_order_xi}
\end{equation}
but now the influence of the $L/16$ $NS$-arcs is accounted in the factor ${\cal P}^{(2)}_{S}(p)$
representing $L/4$ $S$-arcs. As before, we will compute the contributions of the shortest arcs
under the correlations arising from the placement of the $NS$-arcs, and treat the contribution
arising from the longer arcs in a mean-field manner.

In other words, the problem is now reduced to the placement of both $S$ and $NS$ arcs that respect
the constraints imposed by the contact matrix $A$. Obviously, the placement of arcs of one type
introduces additional constraints on the placement of those of other type. First, some of the
places will become forbidden because of the non-crossing constraints. Second, if we are interested
in the complete matching configurations, placing a length-4 arc automatically means placing a
length-2 arc underneath. Therefore, our goal is to place altogether $L/16$ $NS$-arcs, each
\emph{covering} an $S$-arc (in what follows, we will denote a placement of such a construction by
$\blacksquare$), $L/4-L/16=3L/16$ remaining $S$-arcs (as usual, they will be denoted by
$\greysquare$) and $L-4\times L/16-2\times 3L/16=3L/8$ unmatched vertices. The placements are
subject to the non-touching constraints. Proceeding in the same way as in the previous section, we
can write
\begin{equation}
{\cal P}^{(2)}_{S}(p)=\frac{B^{(2)}_{S}(p)}{B^{(2)}_{S}(1)},
\label{eq:bfactors_second_order}
\end{equation}
where the factor $B^{(2)}_{S}(p)$ represents the contributions of the $S$-arcs under the
correlations arising from the presence of the $NS$-arcs. The denominator of the product
\eqref{eq:bfactors_second_order} is given by a multinomial coefficient
\begin{equation}
B^{(2)}_{S}(1) = \frac{\frac{5L}{8}!}{\frac{L}{16}!\frac{3L}{16}!\frac{3L}{8}!}=C^{L/16}_{5L/8}
C^{3L/16}_{9L/16}.
\end{equation}
The multinomial coefficient has the following physical sense: it counts the number of ways to place
length-4 constructions, length-2 arcs and unmatched vertices when all the places are available by
the contact matrix, i.e. to count the number of link configurations of the form
$(\cdots\Circle\Circle\greysquare\Circle\blacksquare\greysquare\Circle\greysquare
\Circle\blacksquare\Circle\Circle\cdots)$. It can be factorized into two binomial coefficients,
describing the following placements: first put $L/16$ length-4 constructions among
$L/16+3L/16+3L/8=5L/8$ objects, and then put $3L/16$ remaining $S$-arcs among $3L/16+3L/8=9L/16$
available places. In what follows, we will evaluate the numerator of
\eqref{eq:bfactors_second_order}.

\subsection{Localization of the problem}

The computation of $B^{(2)}_{S}(p)$ requires counting the number of placements of $S$ and $NS$ arcs
allowed by the contact matrix and respect the non-touching constraint. In order to facilitate the
manipulations, we shall use the following convention: we say that the length-4 construction
(composed of a $NS$-arc which cover $S$-arc underneath) is placed at position $i$ if $i$ is the
starting point of the corresponding $S$-arc, and the whole construction occupies positions
$i-1,i,i+1,i+2$. For simplicity, when discussing the placement of $NS$ arcs, we will also assume
that the corresponding $S$-arc is always placed below. The probability that the $NS$ arc can be
placed at position $i$ is hence given by $p(1-(1-p))=p^{2}$; if the position $i$ is indeed allowed,
it will be denoted by $\boxplus$. Of course, a $S$-arc only can also be placed at position
$\boxplus$. We will mark by $\boxdot$ the positions at which a $S$-arc can be placed, but not the
$NS$-arc (it happens with probability $p(1-p)$). As previously, a position at which none of the
arcs can be placed will be marked as $\square$.

We proceed in a manner similar to what has been done in the section \ref{sec:first-order}. Again,
we notice that the global placement problem on a string of the form
$(\cdots\square\square\boxplus\square\boxdot\square\boxplus\square\boxdot\boxdot\cdots)$ can be
reduced to a set of local CIS problems on independent pieces, separated by a forbidden position
$\square$. These sequences have to be of a special form so that the choice of the placement on one
piece doesn't interfere with the placements on the neighboring pieces.

The optimal form of each independent piece can be shown to be
\begin{equation}
\square\underbrace{\lozenge~\boxed{\whitesquare\cdots\whitesquare}~\lozenge}\square,
\end{equation}
where $\lozenge$ represents a position on which a $NS$-arc is not allowed, i.e. either $\square$ or
$\boxdot$. This requirement is based on the observation that two $NS$-arcs placed in neighboring
sequences have to be separated by at least three non-allowed positions in order to avoid
interference due to the non-touching of the arcs. Same restrictions occur also in the boxed part of
the sequence. One sees that forbidden positions $\square$ must have at least one neighboring
position $\boxplus$ that allow for the placement of the $NS$-arcs; otherwise, the sequence in
question can be separated in two independent pieces according to the definition above.

Each independent piece of length $k$ can be represented by a certain number of different sequences
that satisfy the restrictions below. We illustrate these possible variants in the Table
\ref{tbl:CIS_local_pieces} for different lengths up to $k=3$. The probability of each sequence is
fully determined through the number of positions of different sorts: $k'$ of $\boxplus$, $k''$ of
$\boxdot$ and $k-k'-k''$ of $\square$. The density of each independent sequence is then given by
\begin{equation}
u_{k',k'',k}=p^{2k'}(p(1-p))^{k''}(1-p)^{k-k'-k''}(1-p)^2(1-p^2)^2,
\end{equation}
where the factor $(1-p)^2$ comes from the two forbidden positions $\square$ at the endings of the
sequence, and the factor $(1-p^2)^2$ ensures that the next-to-forbidden position doesn't allow for
the placement of a $NS$-arc, i.e. is $\lozenge$.

\begin{table}[ht]
\begin{centering}
\begin{tabular}{|c|c|c|c|c|c|}
\hline $k$ & $\alpha$ & Representation & $u_{k',k'',k}/((1-p)^2(1-p^2)^2)$ & Non-zero  $Y^{\alpha,k}_{m,n,k',k''}$

\tabularnewline \hline \hline 1 & 1 & $\square\underbrace{\boxdot}\square$ & $p(1-p)$ &
$Y^{1,1}_{0,1,0,1}=1$

\tabularnewline \hline \hline 2 & 1 & $\square\underbrace{\boxdot\boxdot}\square$  & $(p(1-p))^2$ &
$Y^{1,2}_{0,1,0,2}=2$

\tabularnewline \hline \hline 3 & 1 & $\square\underbrace{\boxdot\boxdot\boxdot}\square$ &
$(p(1-p))^3$ & $Y^{1,3}_{0,1,0,3}=3$, $Y^{1,3}_{0,2,0,3}=1$

\tabularnewline \hline 3 & 2 & $\square\underbrace{\boxdot\boxplus\boxdot}\square$ &
$p^{2}(p(1-p))^2$ & $Y^{2,3}_{0,1,1,2}=2$, $Y^{2,3}_{0,2,1,2}=Y^{2,3}_{1,0,1,2}=1$

\tabularnewline \hline 3 & 3 & $\square\underbrace{\boxdot\boxplus\square}\square$ &
$p^{2}(p(1-p))(1-p)$ & $Y^{3,3}_{0,1,1,1}=Y^{3,3}_{1,0,1,1}=1$

\tabularnewline \hline 3 & 4 & $\square\underbrace{\square\boxplus\boxdot}\square$ &
$p^{2}(p(1-p))(1-p)$ & $Y^{4,3}_{0,1,1,1}=Y^{4,3}_{1,0,1,1}=1$

\tabularnewline \hline 3 & 5 & $\square\underbrace{\square\boxplus\square}\square$ & $p^{2}(1-p)^2$
& $Y^{5,3}_{1,0,1,0}=1$

\tabularnewline \hline \hline 4 & 1 & $\square\underbrace{\boxdot\boxdot\boxdot\boxdot}\square$ &
$(p(1-p))^4$ & $Y^{1,4}_{0,1,0,4}=4$, $Y^{1,4}_{0,2,0,4}=3$

\tabularnewline \hline 4 &  & $\cdots$ &  &

\tabularnewline \hline
\end{tabular}
\end{centering}
\caption{The explicit representation of different possible elementary sequences for lengths up to $k=3$.
The local CIS problem is solved independently on each sequence, providing the weights
$Y^{\alpha,k}_{m,n,k',k''}$.}
\label{tbl:CIS_local_pieces}
\end{table}

\subsection{Solution of the CIS problem}

As it has been done in the section \ref{sec:first-order}, we can try to solve the local CIS problem
on these independent sequences. However in the present case the combinatorics is rather involved
since the solution inside each block depends on the distribution of forbidden and allowed
positions. Nevertheless, if we truncate the series at some length $k_{0}$ (as it has been done at
first order of arcs expansion), these solutions can be computed for each sequence via explicit
enumeration, for the pseudo-code see Table \ref{tbl:pseudocode}. Let us call
$Y^{\alpha,k}_{m,n,k',k''}$ a number of ways to put $m$ $NS$-arcs (hiding corresponding $S$ arcs)
and $n$ uncovered $S$-arcs on the sequence of length $k$ comprising $k'$ positions $\boxplus$ and
$k''$ positions $\boxdot$, where $\alpha$ counts the number of different sequences having the same
density $u_{k',k'',k}$.

\begin{table}[h]
\begin{centering}
\begin{tabular}{l}
\hline
set $k_{0}$;
\tabularnewline
set all $Y^{\alpha,k}_{m,n,k',k''}=0$;
\tabularnewline
\textbf{for} $k = 1,\ldots,k_{0}$
\tabularnewline
$\,\,\,\,\,$ \textbf{for} $\alpha = 1,\ldots,\alpha_{max}(k)$
\tabularnewline
$\,\,\,\,\,$ $\,\,\,\,\,$ Generate correct configuration $\alpha$ with $k'$ $\boxplus$, $k''$ $\boxdot$ and $(k-k'-k'')$ $\square$:
\tabularnewline
$\,\,\,\,\,$ $\,\,\,\,\,$ Choose and distribute $k'$ $\boxplus$;
\tabularnewline
$\,\,\,\,\,$ $\,\,\,\,\,$ Neighbors of $\boxplus$ are $\lozenge$, i.e. either $\boxdot$ or $\square$;
\tabularnewline
$\,\,\,\,\,$ $\,\,\,\,\,$ All other elements are $\boxdot$;
\tabularnewline
$\,\,\,\,\,$ $\,\,\,\,\,$ \textbf{for} $m=0,\ldots,k'$
\tabularnewline
$\,\,\,\,\,$ $\,\,\,\,\,$ $\,\,\,\,\,$ \textbf{for} $n=0,\ldots,k'+k''$
\tabularnewline
$\,\,\,\,\,$ $\,\,\,\,\,$ $\,\,\,\,\,$ $\,\,\,\,\,$ Try to place $m$ $NS$-arcs $\blacksquare$ and $n$ $S$-arcs $\greysquare$ on allowed positions;
\tabularnewline
$\,\,\,\,\,$ $\,\,\,\,\,$ $\,\,\,\,\,$ $\,\,\,\,\,$ \textbf{if} non-touching constraints satisfied: increment $Y^{\alpha,k}_{m,n,k',k''}$;
\tabularnewline
$\,\,\,\,\,$ $\,\,\,\,\,$ $\,\,\,\,\,$ \textbf{end}
\tabularnewline
$\,\,\,\,\,$ $\,\,\,\,\,$ \textbf{end}
\tabularnewline
$\,\,\,\,\,$\textbf{end}
\tabularnewline
\textbf{end}
\tabularnewline
\textbf{return} $Y^{\alpha,k}_{m,n,k',k''}$.
\tabularnewline
\hline
\end{tabular}
\end{centering}
\caption{Counting algorithm for the computation of coefficients $Y^{\alpha,k}_{m,n,k',k''}$ up to a
maximum sequence length $k_{0}$.}
\label{tbl:pseudocode}
\end{table}

\begin{figure}[!ht]
\begin{center}
\includegraphics[width=0.57\textwidth]{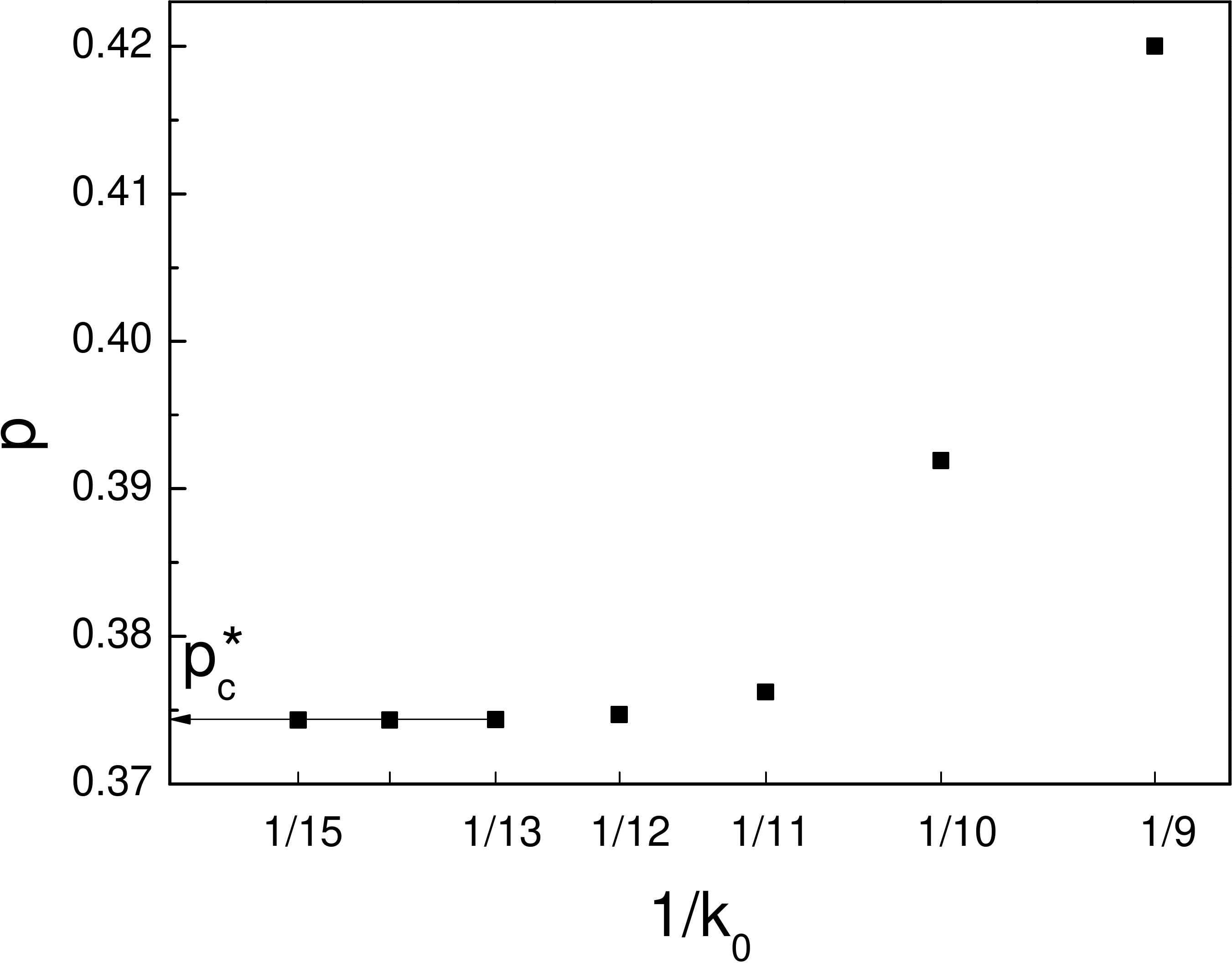}
\end{center}
\caption{Estimation of the prediction $p^{*}_{c}$ for the critical value $p_{c}$ at the second order
of arcs expansion. Each point represents a corresponding estimation when
\eqref{eq:generating_function_second_order} is approximated by a partial product up to the lengths
$k_{0}$. The estimations demonstrate a fast convergence with $k_{0}$ to the value
$p^{*}_{c}=0.3743$.}
\label{fig:second_order}
\end{figure}

Then, again, a generating function for a piece $(k',k'',k)$ can be introduced:
\begin{equation}
W^{\alpha}_{k',k'',k}(s)=\sum_{m}\sum_{n}Y^{\alpha,k}_{m,n,k',k''}s^{m+n},
\end{equation}
where the power of $s$ counts the overall number of $S$-arcs placed at each individual sequence.
Hence the generating function for the whole chain of $L$ a priori available positions reads
\begin{equation}
W(s)=\prod_{k=1}^{k_{0}}\prod_{(k',k'')}
\left(\prod_{\alpha}W^{\alpha}_{k',k'',k}(s)\right)^{Lu_{k',k'',k}}.
\label{eq:generating_function_second_order}
\end{equation}
We want to control that the total number of $S$-arcs is $L/4$, so we are interested in the
coefficient behind the $s^{L/4}$: this will give us precisely the quantity $B^{(2)}_{S}(p)$. This
coefficient is obtained by the integration of $W(s)/s^{L/4}$ around zero:
\begin{equation}
B^{(2)}_{S}(p)=\frac{1}{2\pi i} \oint ds \exp \left[ L( g_{s}(p)-1/4 \log s) \right],
\label{eq:coefficient_integral_length4}
\end{equation}
where $g_{s}(p)=\log{W(s)/L}$. The saddle-point equation is
\begin{equation}
\frac{\partial g_{s}(p)}{\partial s}=\frac{1}{4s}.
\end{equation}
Given the solution of the saddle-point equation $s^{*}$, we have
\begin{equation}
B^{(2)}_{S}(p)=\exp \left[ L( g_{s^{*}}(p)-1/4 \log s^{*}) \right].
\label{eq:coefficient_integral_length4}
\end{equation}
Combining \eqref{eq:bfactors_second_order}, \eqref{eq:second_order_xi} and \eqref{eq:critcond}, and
going in maximum length up to $k_{0}=15$, we get a fast convergence to the value $p_{c}=0.3743$
which is very close to the value found in numerical simulation, see \fig{fig:second_order}.

\subsection{Next orders}

In principle, this estimation can be improved further by considering the contributions from higher
order arcs. For example, for the next order, including length-6, or next-to-next-to-shortest
($NNS$) arcs, we can write
\begin{equation}
\xi^{L/2}(p)= \underbrace{p^{L/4}}_{\rm longer\;arcs}\, \underbrace{{\cal P}^{(3)}_{S}(p)}_{\rm
S-arcs},
\label{eq:third_order_xi},
\end{equation}
where now the $S$-arcs are placed according to the restrictions imposed by $L/32$ $NNS$-arcs and
$L/16$ $NS$-arcs. A subtlety here is that if we are interested in the fully-matched configurations,
each of the $NNS$-arcs may hide either 2 $S$-arcs, either a nested structure of $NS$ and $S$ arcs.
Hence, placing $L/32$ $NNS$-arc on a certain position (allowed with the probability
$p(1-(1-p^{2})^{2})$) automatically means placing $L/64$ $NS$-arcs and $3L/64$ $S$-arcs; at the
same time, $L/16-L/64=3L/64$ of $NS$-arcs are still remaining outside the placed $NNS$-arcs, and
$L/4-3L/64-3L/64=5L/32$ of $S$-arcs are still remaining outside both $NNS$ and $NS$. We have to
place them altogether with $5L/16$ unmatched vertices. Therefore, if we write, as usual,
\begin{equation}
{\cal P}^{(3)}_{S}(p)=\frac{B^{(3)}_{S}(p)}{B^{(3)}_{S}(1)},
\label{eq:bfactors_third_order}
\end{equation}
the factor $B^{(3)}_{S}(1)$ will be given by
\begin{equation}
B^{(3)}_{S}(1) =
\frac{\frac{35L}{64}!}{\frac{L}{32}!\frac{3L}{64}!\frac{5L}{32}!\frac{5L}{16}!}=C^{L/32}_{35L/64}
C^{3L/64}_{33L/64} C^{5L/32}_{15L/32}.
\end{equation}
The calculation of the factor $B^{(3)}_{S}(p)$ would involve, as in the previous sections, the
partitioning of the problem into a set of local CIS problems. The solution to the global problem
could then be obtained by imposing that overall number of $S$-arcs is fixed to $L/4$. Note that, in
principle, we could have thought about fixing the number of longer arcs as well: the number of
$NS$-arcs to $L/16$, the number of $NNS$-arcs to $L/32$, \emph{etc}. This method would require to
introduce several counting variables in the generating function; then we would need to perform the
saddle-point analysis in a multi-dimensional space, which may lead to possible numerical
instabilities. At the same time, the idea to control the number of $S$-arcs, subject to constraints
due to the presence of longer arcs, leads to a saddle-point equation with respect to one variable
only at each order of the expansion, which is easier to control.

\section{Beyond Bernoulli model of planar matching for non-integer alphabets}
\label{sec:beyond-bernoulli}

The planar matching problem has an application to a toy formulation of the optimal secondary
structure problem in RNA molecules. A real RNA is a single-stranded polymer composed of four types of
nucleotides (A, C, G and U). The secondary structure of RNA is represented by the chemical bonds
between the stable Watson-Crick pairs A-U and G-C in the folded state. The simplest theories
designed for the study of statistical properties of the RNA secondary structures usually focus on
the \emph{random} RNAs in which the nucleotide sequence is random, and assume the saturation of
base pairings and the exclusion of the pseudoknots which are known to be rare in real RNAs
\cite{vanBatenburg2000}. These assumptions imply that the secondary structure can be represented as
a planar diagram, the solution of the planar matching problem considered in this paper, where the
contact matrix $A$ encodes the disorder in the primary RNA sequence of nucleotides
\cite{LVTN,BundschuhHwa2002}. In this picture, the parameter $p$ of the Bernoulli contact matrix
$A$ is in the one-to-one correspondence with the number of nucleotides, or alphabet, $c$ (equal to
four for real RNAs) in the primary sequence: since $p$ characterizes the average density of
contacts that each base may have, we simply have $p=1/c$. The contact matrix representation of the
sequence disorder is a convenient tool since it allows to expand the study over the non-integer
alphabets.

Although it is clear that, given the pairing complementarity rules, one can always build a contact
matrix from a given primary sequence, the opposite in general is not true. Indeed, in the Bernoulli
model, each element of the matrix is generated \emph{independently} according to the probability
distribution \eqref{eq:Bernoulli_matrix_probability}, hence, it lacks the transitivity: even if the
elements $A_{ij}$ and $A_{jk}$ appear to be equal to one in the contact matrix, the element
$A_{ik}$ might be zero. However, this limitation is irrelevant in the thermodynamic limit, i.e.
when the length of the sequence $L \to \infty$ \cite{LVTN}.

Still, it would be interesting to understand whether there is a way to construct explicit random
primary sequence that could model the primary sequences with non-integer alphabets. In the context
of the phase transition described in the section \ref{sec:background}, we have seen that there is a
critical value $p_{c}\approx 0.379$ of the bond formation probability that separates the regions of
optimal and non-optimal structures; this critical probability corresponds to the critical alphabet
$c_{cr}\approx 2.64$ in this generalized primary sequence setting. In this section, we address the
following questions: i) Is it possible to construct explicitly a random sequence with transitive or
partially transitive matching rules that would correspond to a non-integer alphabet $c$, i.e. have
a density $p=1/c$ of ones in the contact matrix, generated according to this sequence? ii) Do these
sequences exhibit an analogous critical behavior as the Bernoulli model with the same parameter
$p$, and what is the relation to the behavior of the Bernoulli model?

\subsection{Construction of the non-integer alphabets}

For the models of random sequences, we consider a set of monomers of different types, that we will
call A, B, C, etc. Perhaps the most natural way to think of the non-integer alphabet $2<c<3$ is to
consider three types of monomers: A, B and C, mixed together. For the sake of simplicity, we will
assume that the links can be established between the monomers of the same type, A-A, B-B and C-C.
It is clear that if three types of monomers are distributed randomly and independently along the
sequence, this corresponds to an alphabet $c=3$. However, the effective non-integer alphabet $c<3$
can be modelled by assuming that the distribution of monomers along the chain is correlated.
Suppose that starting from the first randomly chosen monomer, each next monomer in the sequence is
generated according to the Markov-like process, with the probabilities that depend on the monomer
at the previous step:
\begin{center}
\begin{tabular}{c|ccc}
  ~&  A &  B & C  \\
\hline
 A &  $1-2\epsilon$ & $ \epsilon$ & $ \epsilon$ \\
 B&  $\epsilon$ &  $1-2 \epsilon$ & $\epsilon $ \\
 C &  $\epsilon$ & $ \epsilon$ & $1-2\epsilon$  \\
\end{tabular}
\end{center}
This probability matrix is chosen to be symmetric with respect to all monomer types. Each monomer
type appears in subsequences unless it is changed to another type: ($\cdots$A A A B B B B A C C C
$\cdots$). It has been proven in \cite{Vladimirov2013} that if the perfect matching solutions
exist, there is at least one in which the neighboring monomers of the same type are matched
together. It means that without any loss of generality, we can match the repeated monomers along
the chain; this way, each subsequence of a certain type of even or odd length is reduced to one or
zero monomers of this type, respectively: ($\cdots$B A A A C$\cdots)\rightarrow(\cdots$B A
C$\cdots$).

The variation of the parameter $\epsilon$ from $0$ to $1/3$ then gives a sequence that corresponds
to an effective alphabet $c$ in a range from 1 to 3. The relation between $\epsilon$ and $c$ can be
estimated as follows:
\begin{equation}
c=\left( \frac{1}{\epsilon}-2\right)^{2\epsilon} \frac{1}{1-2\epsilon}.
\label{eq:c_epsilon_relation}
\end{equation}
The rational behind this estimation is based on the concept of Shannon information entropy
\cite{Shannon}. The entropy rate of this markovian sequence is given by
\begin{equation}
S = - \sum_{a=\text{A,B,C}} P(a) \sum_{b=\text{A,B,C}} P(b \mid a) \log P(b \mid a),
\label{eq:entropy1}
\end{equation}
where $P(a)=1/3$ is an \emph{a priori} probability for the monomer of a certain type, and $P(b \mid
a)$ is a conditional probability that the monomer of the type $a$ is followed by the monomer of the
type $b$; this probability is given by the probability matrix of the considered Markov process. On
the other hand, if one assumes that the sequences constructed in this way are described by an
effective alphabet with $c$ equivalent monomers, we simple have
\begin{equation}
S = - \sum_{a=1}^{c} \hat{P}(a) \log \hat{P}(a)
\label{eq:entropy2}
\end{equation}
with $\hat{P}(a)=1/c$. The combination of \eqref{eq:entropy1} and \eqref{eq:entropy2} gives us the
relation \eqref{eq:c_epsilon_relation}. Thus constructed alphabet will be referred to as the
``correlated'' alphabet.

\begin{figure}[!ht]
\begin{center}
\includegraphics[width=0.34\textwidth]{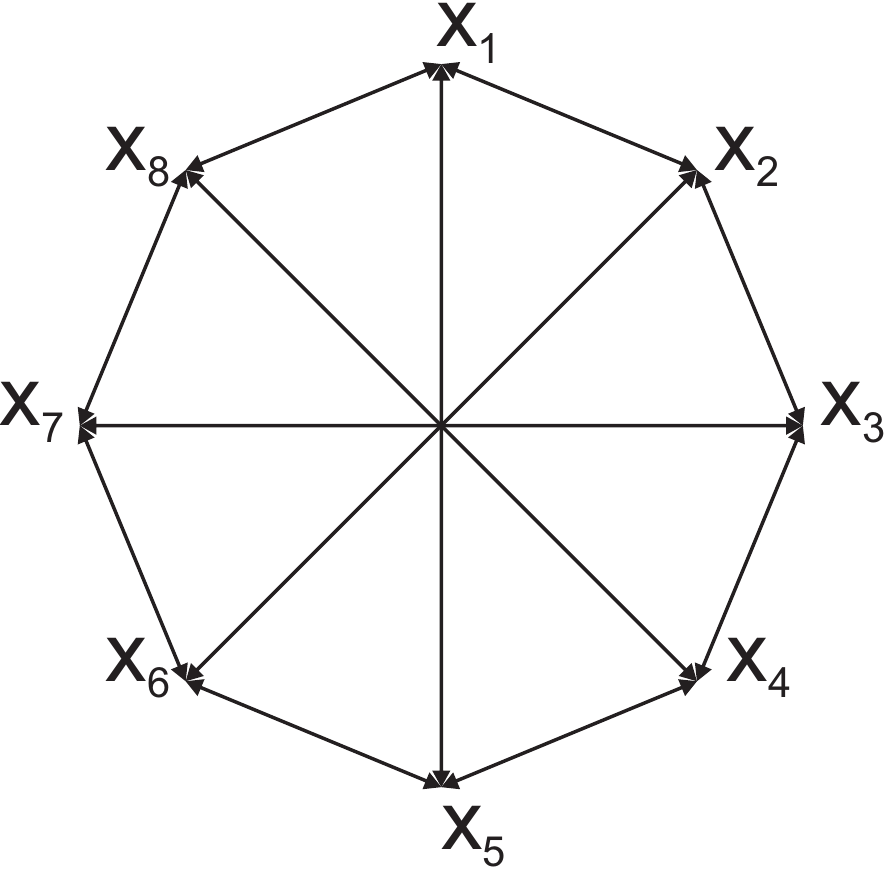}
\end{center}
\caption{An example of the matching rules in the $(8,3)$ rational alphabet model. In this
representation, a link between the monomers of types $X_{i}$ and $X_{j}$ means that they can
potentially form a bond in the matching structure.}
\label{star_alphabet}
\end{figure}

Another model that can be suggested for non-integer alphabets can be obtained using the observation
that each non-integer alphabet $c$ can be approximated by a rational fraction $c=P/Q$. Imagine a
random polymer with $P$ different monomer types $X_{1},\ldots,X_{P}$, but now allow each of them to
bind only with $Q$ other monomer types. The complementary rules can be depicted as a $P$-polygon
with $Q-2$ additional links, where each link means a possible matching between two monomers, see
\fig{star_alphabet} for an example with $P=8$ and $Q=3$. The ``commutation relations'' for the
monomers read
\begin{align}
&\lbrace X_{i},X_{i\pm j}\rbrace = 1 \text{ for $1\leq j\leq[Q/2]$,}
\\
&\lbrace X_{i},X_{i}\rbrace = 1 \text{ if $Q$ and $P$ odd,}
\\
&\lbrace X_{i},X_{i+P/2}\rbrace = 1 \text{ if $Q$ odd and $P$ even,}
\\
&\lbrace X_{i},X_{i+j}\rbrace = 0 \text{ otherwise,}
\end{align}
where $\lbrace X_{i},X_{k}\rbrace$ represents a presence (one) or absence (zero) of possible
matching between the two monomers $X_{i},X_{k}$; the periodic condition $X_{i+P}\equiv X_{i}$ is
understood. We will call this model a $(P,Q)$ ``rational alphabet''. Note that by construction this
alphabet is non-transitive. A particularity of this model is that there is an infinite number of
ways to represent $c$ as a fraction. Let us call $P^{*}$ and $Q^{*}$ as the minimal $P$ and $Q$
that give $c=P/Q$. Then $P=nP^{*}$ and $Q=nQ^{*}$ for an arbitrary integer $n$ give the same value
of $c$, although involving a different number of monomer types. In the thermodynamic limit $L \to
\infty$ it will make no difference since the density of ones in the contact matrix will be exactly
$p=Q/P$, but for finite $L$ it may result in a different behaviors for the models with
$c=P^{*}/Q^{*}$, $c=2P^{*}/2Q^{*}$, etc. In order to minimize this effect, we place ourselves in
the context of the urn model, in which the number of monomers of different sorts in the sequence
are restricted to be equal.

\subsection{Perfect matching transition for non-integer alphabets}

\begin{figure}[!ht]
\begin{center}
\includegraphics[width=0.57\textwidth]{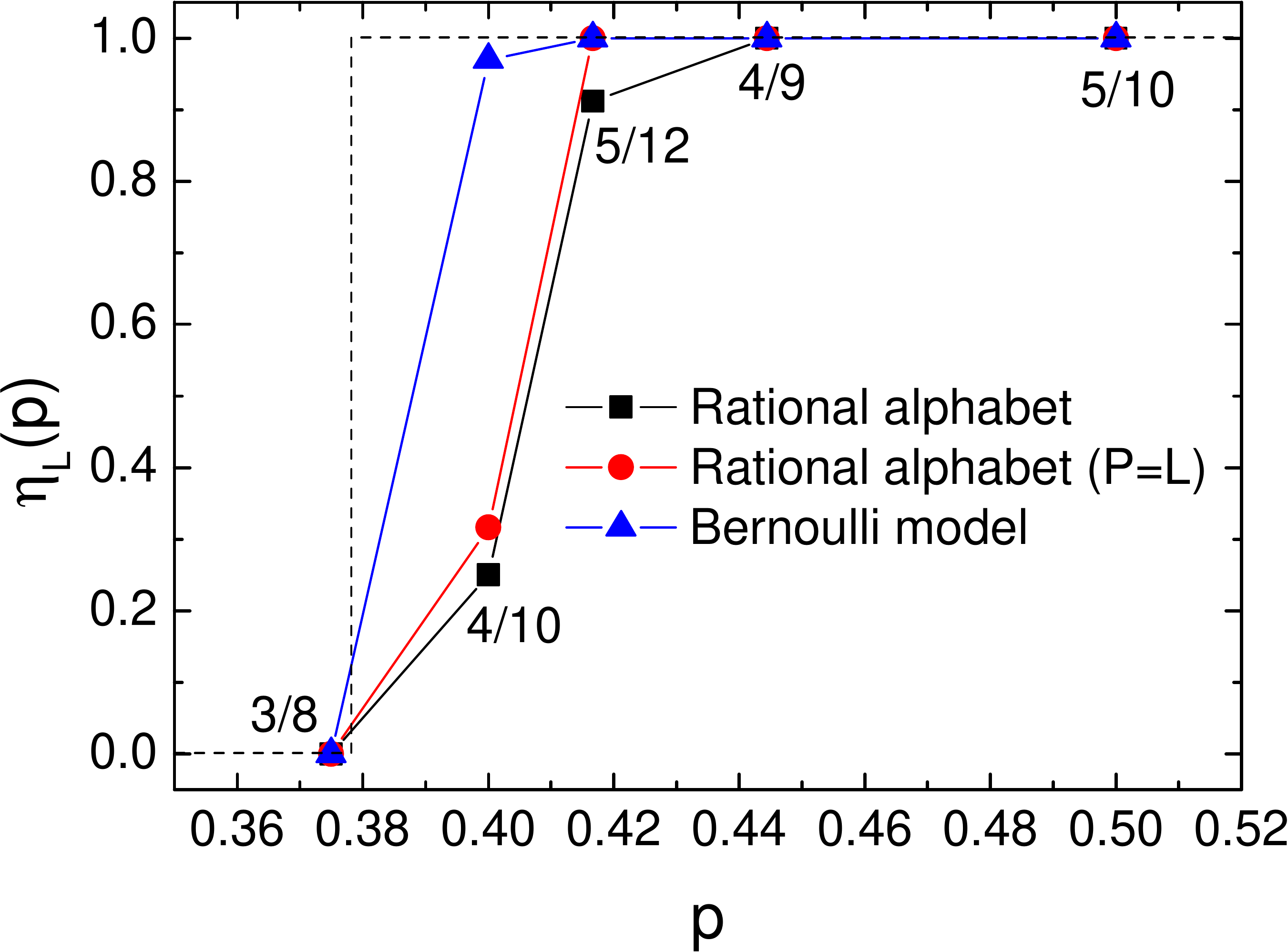}
\end{center}
\caption{The fraction of perfect matchings $\eta_{L}(p)$ as a function of the density $p$ of possible
contacts in the model of $(P,Q)$ rational alphabet (squares, the respective values of $p=P/Q$ are
indicated on the plot), fluctuation-free Bernoulli model (rational alphabet model with $P=L$,
circles) and Bernoulli model (triangles). The simulations have been performed for $L=2000$ and
averaged over 10000 instances.}
\label{transition_alphabets}
\end{figure}

We have investigated the behavior of both correlated and rational alphabets with respect to the
perfect matching transition. To this purpose, we start by drawing random sequences corresponding to
a particular $c$, then construct the contact matrix $A$ according to the matching rules defined in
both models, and, finally solve the matching problem for each instance by the dynamical programming
algorithm.

Surprisingly, we have not observed any transition with variation of $c$ for the model of the
correlated alphabet. In fact, if $c>2$ (or $\epsilon>0.1135$) in this model, there is always a
non-zero fraction of sequences that do not allow for the complete matching solutions. A possible
reason for this is that due to the structure of the sequence, the matching on each subsequence is
easy, but then the sequence is reduced to the primary structure of length $O(L)$ which corresponds
effectively to the alphabet $c=3$, while we know, that for this alphabet (for $p=1/3$) in general
there is no solution to the perfect matching problem.

To the contrary, the rational alphabet model clearly exhibits a critical behavior in the matching
problem. In the \fig{transition_alphabets}, we present the numerical results for the fraction,
$\eta_{L}(p)$, of contact matrices that allow perfect matchings for different $p$. To avoid the
sensitivity on the value of $P$ due to the finite size effects, we have chosen simple test values
$p=Q/P$ with similar $P$ in the range $P \in [8,12]$. The number of perfect matching in these
points are compared to the special case of the limit $P=L$, i.e. when all $L$ randomly distributed
in the chain monomers are distinct, however being able to match $Q=pL$ other monomers in the chain.
This limit corresponds to the fluctuation-free Bernoulli model, in which every line of the matrix
$A$ contains \emph{exactly} $pL$ of ones, without fluctuations of order $\sqrt{L}$ that appear in
the model defined by \eqref{eq:Bernoulli_matrix_probability}. Rational alphabets give similar
predictions, which are however very different with respect to the predictions of the Bernoulli
model. This difference illustrates the ``positive'' role of fluctuations of the number of contacts
in the Bernoulli matrix from the viewpoint of the matching problem.

\section*{Conclusion}

The statistical properties of the planar matching models considered in this paper are fully
determined by only a few parameters: one for Bernoulli model ($p$) and for the model with a
correlated alphabet ($c$, or $\epsilon$), or two for the rational alphabet model ($P$ and $Q$).
Nevertheless, these disordered models exhibit a non-trivial critical behavior. Although an instance of the matching problem can be solved by the dynamical programming algorithm with a
polynomial complexity ($L^{3}$, where $L$ is the size of an instance of the problem), the
analytical estimation of the critical point is hard due to the quenched nature of the disorder.

In this paper, we have developed a combinatorial procedure that allows to obtain successive
estimations for the value of the critical point in the previously studied Bernoulli model. This arcs
expansion procedure benefits from the observation that the arcs of small length play an exceptional
role in the complete matching structures. The key ingredient that makes the problem solvable is the
fact that the global constraint satisfaction problem can be reduced to a set of local ones that are
easier to solve. The developed method hence provides an insight into the fundamental structural
properties of the fully-matched structures.

We have also considered a toy application in the context of random RNA-type sequences. We have designed two simple models that allow for a representation in
terms of a finite set of monomer types and give a concrete sense to the notion of the effective
non-integer alphabet. Although a simple model of a transitive correlated alphabet did not show a
phase transition with a variation of the density of allowed contacts, the non-transitive rational
alphabet clearly manifested the corresponding critical behavior. As a by-product, we have observed
the positive influence of fluctuations in the Bernoulli model by studying the limit of a large
number of monomer types that corresponds to the fluctuations-free case of the Bernoulli model.

Finally, let us emphasize that the models considered in this paper can not be regarded as relevant for the secondary structure formation in \emph{real} RNAs. Indeed, real RNAs are characterized by varying energies of different Watson-Crick parings, limits on minimal lengths of stems (or the so-called stacking energies) and loops, presence of pseudoknots, \emph{etc}. Therefore, the obtained exact quantitative results are certainly not directly applicable to a real RNA. However, as it has been argued in \cite{ValbaTammNechaev2012}, the considered morphological transition is a universal phenomenon, persisting as well in more detailed models of RNA secondary structure formation: it represents a transition from a highly degenerate nearly-perfect structure for small nucleotide alphabets to the unique, but highly defective imperfect structure for large alphabets. We anticipate that the techniques developed in this paper will be useful for the analysis of the transition of this type in more general models.

\section*{Acknowledgments}

The authors are thankful to C. Moore and V. Stadnichuk for fruitful conversations. This work is
partially supported by ``Investissements d'Avenir'' LabEx PALM (ANR-10-LABX-0039-PALM) project PRONET, and the IRSES projects FP7-PEOPLE-2010-IRSES 269139 DCP-PhysBio and
FP7-PEOPLE-2014-IRSES 612707 Dionicos. O.V.V, S.K.N. and M.V.T. are grateful to the Higher School
of Economics program for basic research.

\end{document}